%% file: MAIN_ASE_NIER.tex
\documentclass[sigconf]{acmart}

\AtBeginDocument{%
  \providecommand\BibTeX{{%
    \normalfont B\kern-0.5em{\scshape i\kern-0.25em b}\kern-0.8em\TeX}}}

\begin{document}

\title{End-to-End Rationale Reconstruction} 
\author{Mouna Dhaouadi}
\orcid{0000-0001-9336-7714}
\affiliation{%
  \institution{DIRO, Universit{\'e} de Montr{\'e}al}
  \city{Montreal}
  \country{Canada}
}
\email{mouna.dhaouadi@umontreal.ca}

\author{Bentley James Oakes}
\orcid{}
\affiliation{%
  \institution{DIRO, Universit{\'e} de Montr{\'e}al}
  \city{Montreal}
  \country{Canada}
}
\email{bentley.oakes@umontreal.ca}

\author{Michalis Famelis}
\orcid{}
\affiliation{%
  \institution{DIRO, Universit{\'e} de Montr{\'e}al}
  \city{Montreal}
  \country{Canada}
}
\email{famelis@iro.umontreal.ca}

\renewcommand{\shortauthors}{Dhaouadi, et al.}

\begin{abstract}
\input{text/abstract}

\end{abstract}

\keywords{Software rationale, Natural Language Processing}

\maketitle

\newcommand\mouna[1]{\textcolor{green}{#1}}
\newcommand\todo[1]{\textcolor{red}{#1}}
\newcommand\bentley[1]{\textcolor{blue}{#1}}

\input{text/introduction}


\input{text/example}

\input{text/proposal}
\input{text/graph_consistency_checking}



\input{text/relatedWork}

\input{text/conclusion}


\newpage

\balance

\bibliographystyle{ACM-Reference-Format}
\bibliography{ase_nier}

\end{document}

%% file: text/abstract.tex
The logic behind design decisions, called design \textit{rationale}, is very valuable. In the past, researchers have tried to automatically extract and exploit this information, but prior techniques are only applicable to specific contexts and there is insufficient progress on an end-to-end rationale information extraction pipeline. Here we outline a path towards such a pipeline that leverages several Machine Learning (ML) and Natural Language Processing (NLP) techniques. Our proposed context-independent approach, called \textit{Kantara}, produces a knowledge graph representation of decisions and of their rationales, which considers their historical evolution and traceability. We also propose validation mechanisms to ensure the correctness of the extracted information and the coherence of the development process. We conducted a preliminary evaluation of our proposed approach on a small example sourced from the Linux Kernel, which shows promising results.


%% file: text/introduction.tex
\section{Introduction}
\label{sec:intro}

Software development is a process that requires making several decisions.  The logic behind these decisions, or the \textit{rationale}, represents beneficial information as it can serve as a valuable repository of past experiences that could be used to solve similar problems, and to learn best practices \cite{burge2008rationale}. Thus, many researchers \cite{lee1991extending,noble1988issue,burge2005software,conklin1988gibis} have tried to capture and represent this logic in order to exploit it. In fact, making rationale available in a well-structured and exploitable format will better support decision-making \cite{liang2012learning}, and would foster collaboration, coordination, and project management \cite{burge2008rationale}. But this line of research has not seen wide adoption because of the large effort required to capture the rationale  manually (i.e. the \textit{capture problem}), and of its inability to provide immediate results (i.e. \textit{the cost-benefit problem}) \cite{burge2008design}. The solution is thus to extract and structure rationale automatically,  in a non-intrusive way and with no extra-effort, solving both the \textit{capture} and the \textit{cost-benefit} problems. However, this is a difficult task, as rationale appears in an implicit, poorly organized way and scattered across different artifacts \cite{brunet2014developers}.

Recently, several researchers tried to extract rationale automatically with Natural Language Processing (NLP) techniques \cite{rogers2012exploring,lester2020using,kleebaum2021continuous,sharma2021extracting}. However, their techniques are only applicable to specific contexts and, to the best of our knowledge, there are no techniques allowing their adaptation to other contexts, which hinders their adoption in practice. Additionally, there is no prior work that has proposed a complete end-to-end pipeline that encompasses the extraction, the structuring and the management of rationale, which limits their usefulness. Finally, to the best of our knowledge, none of the previous work have considered capturing the temporal evolution of decisions and of their rationales.

Our vision is therefore to \textit{assist decision makers, whether they are engineers, developers, or others, in any context, with an automated design rationale extraction and management system}. Specifically, we propose to adopt the community vision of an automated \textit{on-demand developer documentation (OD3)} system \cite{robillard2017demand}. An OD3 system has two components: an 
\textit{information inference} component that extracts and structures the useful distributed information, and a \textit{response generation} component that produces responses to the users queries.

This paper presents our initial ideas about how to build the \textit{information inference} component, that we call ``\textit{Kantara}''. The \textit{Kantara} approach aims to automatically extract the decisions and their rationales, from different textual sources and artifacts, then to structure them as a graph-based representation. 
We thus assume that written artifacts produced by stakeholders during their communications, or while executing their tasks, contain rationale information.
We propose to leverage advances in NLP and Machine Learning (ML) to create a pipeline for rationale reconstruction \cite{yue2018design}. The resulting knowledge graph can then be used to a) understand the historical development process of the decisions, b) get a shared understanding of the logic behind the decisions made across both time and space, and c) ensure the consistency and the coherence of the development process by looking for conflicts and by detecting whether a proposed decision interferes with previously agreed upon decisions -- or merely repeats what was tried in the past. In the future, we envision the second component of our envisioned \textit{OD3} system to be a bot that leverages the knowledge graph created by \textit{Kantara}.

\textit{Contributions and Structure.} This paper is structured as follows: i) In Section~\ref{sec:example} we give a running example sourced from the Linux Kernel. ii) The associated \textit{Rationale and Decision Graph} for this example is presented in Section~\ref{sec:rat_dec_graph}. This is a knowledge graph-based representation of design decisions and rationales that takes their historical evolution into consideration and that provides traceability to artifacts. iii) In Sections~\ref{sec:kantara} and~\ref{sec:eval} we present and discuss the \textit{Kantara} approach: an end-to-end generic NLP-based information extraction pipeline to extract and structure design rationale in a knowledge graph. iv) In Section~\ref{sec:cons_checking} we propose  NLP-based mechanisms to detect inconsistencies in the knowledge graph. 
We survey related work in Section~\ref{sec:related} and conclude in Section~\ref{sec:conclusion}.


%% file: text/example.tex
\section{Motivating Example}
\label{sec:example}






\begin{table*}[t]
    \centering
    \footnotesize
    \caption{Commits about reclamation of memory in the OOM Killer.}
    \label{tab:linux_commits}
    \begin{tabular}{p{4.3cm} | c | p{9.6cm}}
    \textbf{Summary} & \textbf{Date} & \textbf{Description}\\\hline
      \texttt{oom: give the dying task a higher priority}   & Aug 9th, 2010 & Increase the dying task's priority to the lowest real-time level such that it is scheduled sooner and can free its memory\\\hline
      
      \texttt{memcg: give current access to memory reserves if it's trying to die}   & Mar 23rd, 2011 & Within a \textit{control group} of threads, increase the priority of a dying task,\\\hline
      
      \texttt{oom-kill:} \newline \texttt{remove boost\_dying\_task\_prio()} & Apr 14th, 2011 & Threads with real-time priorities placed within the \textit{cpu control group} could never run, causing the kernel to hang. Previous commits were reverted. \\\hline
      
      \texttt{mm, oom: introduce oom reaper} & Mar 25th, 2016 & A kernel thread is used for recovering memory (\textit{reaping}) from dying threads, to improve reliability$^{\mathrm{a}}$. \\\hline
      
      \texttt{mm: introduce process\_mrelease system call} & Sep 2nd, 2021 & Introduce a mechanism such that a user process (such as the Android process manager) can request that the memory of a dying thread is reaped$^{\mathrm{b}}$.\\\hline
      \multicolumn{3}{l}{$^{\mathrm{a}}$ \url{https://lwn.net/Articles/668126/}~~ $^{\mathrm{b}}$ \url{https://lwn.net/Articles/864184/}}\\
    \end{tabular}
\end{table*}

We have selected the Out-Of-Memory Killer (OOM-Killer) component in the Linux kernel and its related commits/discussions as the running example and the object of our preliminary evaluation. The OOM-Killer is a last-ditch effort by the kernel to free up memory when tasks have requested all available memory. That is, when there is not enough system memory then future requests for memory will fail and the system may crash. To prevent this, the OOM-Killer will a) select a task to kill, then b) signal (or force) that task (the \textit{OOM victim}) to release its memory and exit. This has been a controversial component from its first suggestion in 1998\footnote{\url{https://marc.info/?l=linux-kernel&m=90222140830612&w=2}}, as some developers do not agree with the strategy that a task is killed without the user's intervention. Nevertheless, this component can be useful to return a system to stability.

We chose the OOM-Killer component for three reasons. First, it has a well-defined scope which is easy to comprehend yet semantically  rich. Second, it has a number of heuristics and interesting decisions, such as the assigning of a \textit{badness score} when deciding which task to kill and the mechanisms for reclaiming memory from the OOM victim. Finally, it is small in terms of commits and code\footnote{The primary file \texttt{mm/oom\_kill.c} is 1205 lines of code, with 410 commits since the start of the Linux kernel Git repository in 2005.}. For the presentation and the initial evaluation of our approach, we manually selected interesting commits from the Git history of OOM-Killer as our example decisions. Specifically, the commits are on the topic of \textit{reclaiming used memory from the OOM victim}.

In Linux kernel development, patches must contain a description of the motivation/rationale behind them. 
Traceability information is provided in multiple ways, such as a) patches are encouraged to add explicit links to the Linux Kernel Mailing List (LKML) discussions in their descriptions, and b) patches must have a \textit{summary phrase} (e.g., ``oom: give the dying task a higher priority''). 
Thus, the kernel and the LKML form a comprehensive repository of decision/rationale information for each commit.

Table~\ref{tab:linux_commits} presents the details for the five relevant OOM-Killer commits, including the \textit{summary phrase}, the \textit{commit date}, and our \textit{description} of the commit. The first two commits in Table~\ref{tab:linux_commits} increase the priority of dying tasks so they can quickly release their memory. The third commit reverts the first two, as conflicts were found between the priority escalation and another mechanism in the kernel. The fourth commit is an approach where a dedicated thread performs the memory reclamation, avoiding the priority escalation issues. The last commit implements a system call such that a user process can ask the kernel to reclaim the memory of a dying task. 

%% file: text/proposal.tex
\section{End-to-End Rationale Recovery}
\label{sec:proposal}

In this section, we present and discuss the proposed \textit{Kantara} approach to automatically create a \textit{Rationale and Decision Graph} from the Linux-sourced example commits. Note that \textit{Kantara} is not Linux-specific, as any commit messages are sufficient for our approach. This example simply provides a best-case traceability scenario.


\subsection{Rationale and Decision Graph}
\label{sec:rat_dec_graph}

\begin{figure*}[t]
\centerline{\includegraphics[width=\textwidth]{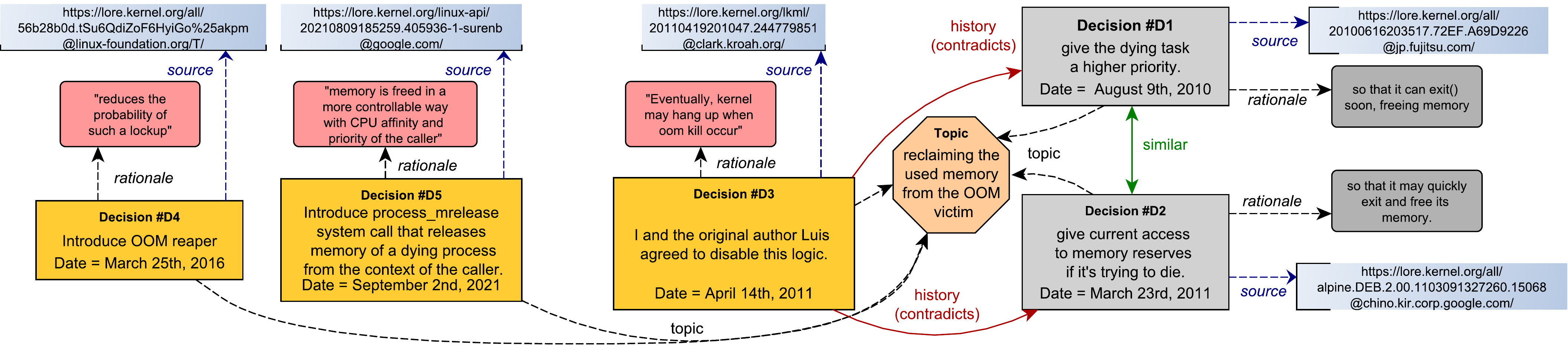}}
\caption{The \textit{Rationale and Decision Graph} for the commits in Table~\ref{tab:linux_commits}. 
}
\label{fig:linux_graph}
\end{figure*}

A \textit{Rationale and Decision Graph} represents design decisions and the rationales behind them as nodes, and the relationships between those entities as edges. For example, the decisions and their rationale for the commits in Table~\ref{tab:linux_commits} are represented by the graph in Fig.~\ref{fig:linux_graph}.  Each decision has an associated \textit{source} element 
(the URI of the artifact from which the decision has been extracted)  to ensure traceability, and is related to a specific \textit{topic} element. Topics organize the extracted decisions into clusters so that when a developer wants to know about a specific design topic, they can easily check all its associated decisions. For instance, anyone working on the \textit{memory reclamation} topic should be pointed to the failed approach (D1/D2/D3), the killer kernel thread (D4), and the system call (D5). 

The main components of our knowledge graph are the triples \textit{(decision, ``rationale'', rationale)}. 
For instance, the decision D1: \textit{give the dying task a higher priority} has the associated rationale:  \textit{so that it can exit() soon, freeing memory}. 
Since rationale is very subjective, engineers can formulate it differently \cite{robillard2016disseminating}. Thus, our graph may be very diverse in the content of its rationale nodes.

The graph also introduces different relationship types between decisions: 1) a \textit{history} relationship that means that a decision is an evolution of another (D3/D1), 2) a \textit{similar} relationship that means that the decisions are semantically similar (D2/D1), and 3) a \textit{contradicts} relationship that means that the decisions might be in conflict (D3/D5). The \textit{contradicts} relationship can also be combined with the \textit{history} one, in the sense that the new decision contradicts its previous version (D3/D1, D3/D2). This captures the fact that the third decision D3 (taken in the third commit) reverts the first two D1/D2 (taken in the first two commits) as explained in Section~\ref{sec:example}.

The goal of the \textit{Rationale and Decision Graph} representation  is to help stakeholders develop a shared understanding of how decisions are interconnected to counter \textit{design evaporation}~\cite{robillard2016sustainable}. Our graph does not only capture the reason behind a decision, but also  contextualizes it in time, which helps recover the holistic view of the development process and thus provides stakeholders with the big picture and helps avoid wasted effort. This representation is very simple and generic and can easily be translated to more specific rationale metamodels, such as IBIS~\cite{conklin1988gibis}, which makes it easily adaptable for different contexts.

\subsection{Approach Overview}
\label{sec:kantara}

Here we describe and discuss the \textit{Kantara} approach to automatically create such a \textit{Rationale and Decision Graph} using ML and NLP. Given textual artifacts, our approach aims to extract the design decisions taken, their historical evolution, their corresponding rationales, and how these decisions are interconnected. We envision \textit{Kantara} to be context-independent (as we will detail later) for broader applicability. The \textit{Kantara} information extraction pipeline has two stages described below: 1) \textit{extracting decision-rationale triples}, and 2) \textit{extracting decision-decision relationships}.
\subsubsection{Extracting Decision-Rationale Triples}\hfill

\textit{Decisions extraction.} We plan to identify decision sentences by classification. For the example in Fig.~\ref{fig:linux_graph}, we redeveloped the decisions classifier from \cite{kleebaum2021continuous} and used it to label the decision-containing sentences in the different commits in Table~\ref{tab:linux_commits}. The classifier correctly labeled D3, D4, and D5 as relevant sentences for decisions. 
However, it incorrectly labeled D1 and D2, as non-relevant decision sentences. The \textit{source} element contains the link to the mail from which the decision has been extracted. 

Since our goal is to offer a context-independent extraction system, our next steps include building an improved and generic decisions classifier. Previous work have proposed context-specific design decisions classifiers~\cite{kleebaum2021continuous, bhat2017automatic, li2020automatic}. However, recent research~\cite{mahadi2020cross} showed that these systems do not offer \textit{conclusion stability} (i.e. the learners are unable to transfer to other contexts). To overcome this, Mahadi et al.~\cite{mahadi2022conclusion} created \textit{software engineering word embeddings}, and found that they improved the classifiers performance. Thus, in our next steps, we plan to reuse these available embeddings. Furthermore, the classifier performance depends alot on its training data. To mitigate this potential limitation, we could label more data, use automatic labeling approaches, or enhance the classifier's performance by combining it with a heuristic-based approach.

\textit{Rationale extraction.} Once we have the decision-containing sentence, we try to extract the reason behind the decision at the sentence level using Semantic Role Labeling (SRL)~\cite{palmer2010semantic}. This NLP technique is able to extract the cause or the purpose of a certain event, based on the grammatical construction of a sentence, so it is context-independent. In Fig.~\ref{fig:linux_graph}, SRL successfully extracted the rationales behind D1 and D2. 
However, since it works at the sentence level, it would fail if rationale is scattered over multiple sentences (as it would capture incomplete rationale), or if the rationale is contained in a different sentence (e.g. D3). We try to mitigate this limitation with the \textit{source} elements so developers can double check the extracted rationale from the original source. Our future plan include  investigating  paragraph-level rationale extraction approaches.

For D4 and D5, our observations have led us to consider investigating using the \textit{manner} information as potential useful information for rationale extraction. 
For instance, in the case of D5 the rationale is found in the sentence:  \textit{"This way the memory is freed in a more controllable way with CPU affinity and priority of the caller."}, and is phrased as the \textit{``how?''} (i.e. \textit{"in a more controllable ..."}). The SRL labeled that sentence part as \textit{manner} information. In the future, we intend to investigate this particular way of phrasing rationale.

Finally, we note that our approach only recovers \textit{recorded} rationale (albeit from sources that were not originally intended to be rationale documents), and not the whole \textit{actual} rationale as that is often never recorded, even unintentionally. This limits its usefulness, and could even have negative effects if wrong decisions were taken based on partial rationale~\cite{horner2006effective}. 

\subsubsection{Extracting Decision-Decision Relationships} \hfill

Once the decision/rationale nodes are created, our pipeline tries to label the relationships between decisions in a \textit{context-independent} way. These relationships are  
the basis of our validation mechanisms.

\textit{Relatedness relationship (Topic).}
Topics are introduced as means to capture relatedness between decisions.
Our initial idea is training a ML classifier on a related/non-related sentences dataset. We obtained this dataset after considering \textit{indirect}, \textit{direct} and \textit{duplicate} links as \textit{related}, and \textit{isolated} links as \textit{non-related}, for entries where at least one of the sentences has the \textit{design} tag from the StackOverflow (SO) relatedness dataset~\cite{shirani2019question}. Since this dataset contains 300K entries spanning the entire SO community,  we argue that our resulting classifier is context-independent. The classifier performed poorly on our example because of imbalanced classes, thus, we plan to investigate the use of techniques such as SMOTE~\cite{fernandez2018smote}. We plan to investigate using keyphrase generation systems~\cite{meng2017deep} to automatically generating the topic title (such as \textit{``reclaiming the used memory from the OOM victim''}).

\textit{Similar relationship.} There are many off-the-shelf approaches for detecting semantic similarity.
The edge between D1/D2 was added after detecting a similarity of 0.86 between them using SpaCy~\cite{vasiliev2020natural}.

\textit{History relationship.} For two related decisions (i.e. ones with the same topic), we check whether one is a previous version of the other. 
We plan on detecting this historical information using heuristics (e.g. the decisions concern the same files, they were done by the same people, one of them refers to the other,  timestamps, etc). For instance, D3 happened after D1, the mail containing D3 referred to D1 and the author of D1 ``acked'' (`acknowledged') D3. Identifying the right heuristics requires extensive empirical evaluation.

\textit{Contradicts relationship.} We propose to capture this relationship using the \textit{contradiction} label of the  Natural Language Inference (NLI) task~\cite{maccartney2009natural}. We fine-tuned the BERT model~\cite{devlin2018bert} on SNLI corpus~\cite{bowman2015snli}. Although the resulting model worked well in other cases (e.g. it detected a contradiction of 0.94 between \textit{"We need to implement this feature to be able to satisfy the requirements"} and 
\textit{"There is no need to do anymore changes"}),  it performed poorly on our example (i.e. D3/D1 and D3/D2) because the contradiction is not obvious. We plan to mitigate this by introducing heuristics (e.g. if the commit of the decision is a revert of another commit, or to exploit the existence of keywords such as \textit{revert} or \textit{disable}).

\subsection{Preliminary Evaluation Summary}
\label{sec:eval}

Table \ref{tab:evaluation} summarizes the evaluation results of the different steps of the \textit{Kantara} pipeline as discussed throughout Section~\ref{sec:kantara}. These promising results prompt us to carry on our work.

\begin{table}[h]
  \caption{Evaluation Summary}
  \label{tab:evaluation}
  \footnotesize
  \begin{tabular}{cl}
    \textbf{Step in the pipeline} &\textbf{Preliminary Evaluation Results} \\
    \midrule
    \textit{Decisions extraction}  & partially successful \\
    \textit{Rationale extraction} & partially successful  \\
    \textit{Relatedness relationship (Topic)} & unsuccessful \\
    \textit{Similar relationship} & successful  \\
    \textit{History relationship} & not applicable (future work) \\
    \textit{Contradicts relationship} & partially successful \\
  \bottomrule
\end{tabular}
\end{table}

\balance

%% file: text/graph_consistency_checking.tex
\section{Graph Validation Mechanisms}
\label{sec:cons_checking}

The validation mechanisms aim to a) verify the consistency of the extracted graph and of the development process, and b) look for potential conflicts when a new decision is proposed.

To satisfy the first goal, we propose a mechanism for detecting inconsistencies between the rationales of the  decisions, and the decisions themselves. For instance, decisions D1 and D2 have a \textit{similar} relationship. Our mechanism should thus verify the relationship between their rationales. We computed a semantic similarity of 0.92, which is a strong indicator of consistency. Moreover, this high value suggests incorporating \textit{duplicate detection} in our future work, as it could be useful to detect the case when the same rationale has resulted in different decisions.  If our mechanisms have found a contradiction between the rationales, then this would have indicated a problem in the graph construction, or even a deeper problem in the reasoning.

To satisfy the second goal, we propose a verification mechanism that should happen at the introduction of every new decision. Let's consider Suren, a developer of the Memory Management (\textit{mm}) sub-system  in the Linux kernel. In 2021, Suren proposes to introduce a process (commit 5 in Table~\ref{tab:linux_commits}) to reclaim the memory of a dying task. At the introduction of D5, our mechanism should detect a  \textit{similar} relationship between D5 and D1 (we computed a semantic similarity of 0.87), and considering the graph structure (i.e. D2 is similar to D1, and D3 contradicts D1 and D2),  this should inform the developer that a similar approach has been tried out more than a decade ago regarding the Out-Of-Memory (oom) component (D1/D2), that it has been abandoned after a short while because of a corner case identified in 2011 (D3), and that their proposed decision (D5) may cause conflict with this previously made decision (i.e. D3). Thus, the verification mechanism would help stakeholders avoid collisions and make the right design choices, which is particularly important in the case of long-term, geographically-distributed or large-scale projects. Consequently, this will result in a better design quality and would help solve the \textit{design erosion} problem \cite{van2002design}. In this particular case, we should note that Suren was aware of the earlier approaches because he discussed the commit logs with the maintainers on the LKML. In fact, the proposed approach in D5 introduces a \textit{system call}, which avoids the priority escalation problem of D1/D2.


%% file: text/relatedWork.tex
\section{Related Work}
\label{sec:related}

In \cite{liang2012learning}, the authors focus on the design of an algorithm to extract and structure  design rationale from design documents based on the 
ISAL model \cite{liu2010new}. They use artifact information identification, issue
summarization and solution–reason pair discovery,  and leverage semantic graph-based algorithms. Unlike our work, they do not consider the historical evolution of decisions. In~\cite{viviani2019locating}, the authors attempted to recover design information from developer discussions. Similar to our work, they try to extract design information; however they do not attempt to structure it.
In \cite{rastkar2013did}, the authors tried to generate concise descriptions of the reasons resulting in a code change. To do so, they leveraged advanced summarization techniques. Thus, they mainly differ from us in the used approach, and they only consider rationale in the context of code change. 
In \cite{mccall2018using}, the authors propose ASGAR, a semantic grammar-based approach to automatically capture and structure design rationale. This work differs from ours by the use of  semantic grammars.


%% file: text/conclusion.tex
\section{Conclusion}
\label{sec:conclusion}

In this paper, we present our initial ideas for an automated end-to-end graph-based rationale reconstruction approach that we call \textit{Kantara}. A small example of commits for the Out-Of-Memory Killer of the Linux kernel are used to illustrate and evaluate the two main steps of 1) \textit{extracting decision-rationale triples}, and 2) \textit{extracting decision-decision relationships}. 
We also explain our proposed  mechanisms to ensure the consistency of the graph.
Preliminary evaluation results indicate promise, and we describe our plans concerning the observed limitations. 
To evaluate the usefulness of our approach, we are now building the response generation component of our envisioned OD3 system. A prototype implementation as a GitHub bot can then assist end-users directly during the development process. 
